\documentclass[twocolumn]{aastex631}

\usepackage{hyperref}

\begin{document}

\title{Is Earendel a Star Cluster?: Metal Poor Globular Cluster Progenitors at $z\sim6$}

\author[0000-0002-2282-8795]{Massimo Pascale}
\affiliation{Department of Astronomy, University of California, 501 Campbell Hall \#3411, Berkeley, CA 94720, USA}

\author[0000-0003-2091-8946]{Liang Dai}
\affiliation{Department of Physics, University of California, 366 Physics North MC 7300, Berkeley, CA. 94720, USA}

\author[0000-0003-1625-8009]{Brenda L.~Frye}
\affiliation{Department of Astronomy/Steward Observatory, University of Arizona, 933 N. Cherry Avenue, Tucson, AZ 85721, USA}

\author[0000-0002-9861-4515]{Aliza G. Beverage}
\affiliation{Department of Astronomy, University of California, 501 Campbell Hall \#3411, Berkeley, CA 94720, USA}

\begin{abstract}

The strongly-lensed $z\sim 6$ Sunrise galaxy offers an incredible opportunity to investigate star formation in the early universe on parsec or smaller scales. The highly magnified object Earendel within the Sunrise was previously identified as a candidate star or binary due to size constraints placed by the lensing magnification, however recent works have suggested this constraint may be relaxed to even the size of star clusters. Here, we explore the hypothesis that Earendel may actually be a star cluster, and simultaneously evaluate other star clusters within the host galaxy. Leveraging deep, archival James Webb Space Telescope NIRSpec PRISM spectroscopy, we determine a spectroscopic redshift for the Sunrise galaxy $z=5.926 \pm 0.013$, and we fit simple stellar population (SSP) models from three premier libraries to evaluate the physical parameters of Earendel and another distinct star cluster in the Sunrise dubbed `$1b$'. We find the rest-UV through optical continuum of Earendel to be well-described by an SSP, nearly equivalently to $1b$ which is confidently a star cluster. We infer they have intermediate ages $t_{\rm age}\sim 30$--$150\,$Myr, are metal poor ($Z_\star\lesssim10\%\,Z_\odot$), and are consistent with the formation age-metallicity trend seen in local globular clusters. Such intermediate age clusters are seldom probed spectroscopically in the high redshift universe, and we explore the extent to which these clusters can be characterized via the spectroscopic continuum.

\end{abstract}

\section{Introduction} \label{sec:intro}

Strong gravitational lensing has enabled the study of distant galaxies at unprecedented spatial scales \citep[e.g.,][]{Vanzella2017protoGCs,Mowla2022}. High magnification from foreground galaxy clusters stretches  background galaxies, resolving individual stellar associations such as parsec-scale star clusters at a high signal-to-noise ratio \citep[][]{Pascale2023, Vanzella2023}. While strong-lensing magnification is typically on the order of dozens or fewer, compact stellar sources serendipitously situated near the lensing caustic can be magnified by factors of hundreds to thousands \citep[e.g.,][]{Welch2022a, Pascale2024}. In extreme cases, even individual stars may be detectable in sensitive space-based imaging \citep{MiraldaEscude1991,Windhorst2018,Dai2018Abell370,Diego2019ExtremeMagnification}. These have been observed in low to intermediate ($z\lesssim1$) galaxies strongly-lensed by a foreground galaxy cluster~\citep[e.g.,][]{Kelly2018Icarus, Rodney2018NatAsM0416Transients, Chen2019, Kaurov2019, Pascale2022, Fudamoto2024}, assisted by temporary flux boosts induced by microlensing from intracluster stars~\citep{Venumadhav2017CausticMicrolensing, Diego2018DMUnderMicroscope, Oguri2018CausticMicrolensing, Weisenbach2024MLnearMacroCausticsReview}.

In other cases, it is thought the source-caustic alignment can be so fortuitous that strong-lensing alone delivers sufficiently large magnification factors ($\sim 10^4$) to reveal individual stars in high redshift galaxies ($z\gtrsim5$) in deep Hubble Space Telescope ({\it HST}) or James Webb Space Telescope ({\it JWST}) imaging \citep[e.g.,][]{Meena2023,Furtak2024,Diego2023mothra}. The first such example reported is Earendel \citep{Welch2022a,Welch2022b} at $z\sim6$, which remains the highest redshift lensed star candidate identified to date. Earendel, or WHL 0137-LS, was discovered in the WHL J013719.8-82841 galaxy cluster field in {\it HST} imaging acquired through the RELICS program \citep{Coe2019, Salmon2020}. The {\it HST} discovery was first presented in \cite{Welch2022a}, and later with {\it JWST} NIRCam imaging in \cite{Welch2022b}. Earendel is thought to be persistently magnified by factors of $10^3-10^4$ due to a close proximity to the lensing critical curve, and that the image flux appears stable across a two-year baseline in the observer frame, indicating that its brightness is not primarily a result of microlensing. 

Synthesizing the predictions of four strong-lensing models of the foreground galaxy cluster ($z_{\rm cluster}$=0.566), \cite{Welch2022b} constrain the intrinsic size of Earendel to be $r\lesssim 0.02$~pc, or $\sim 4000$~AU, placing it at the scale of an individual stellar system rather than a star cluster. \cite{Welch2022b} found {\it JWST} NIRCam photometry is not well fit by a single star, but rather that a binary system of two $\sim 20\,M_\odot$ stars of effective temperatures $\sim 34000$~K and $9000$~K is more consistent with the data. Given the tight size constraint provided by strong-lensing models, a simple stellar population (SSP) fit has not been explored in that work or other prior works on Earendel.

\cite{JiDai2024} recently identified that in cases such as Earendel, deriving size constraints solely from macro strong-lensing models can be prone to bias, as such models only include dark matter subhalos which have galaxies visible in {\it HST} or {\it JWST} imaging. In the hierarchical picture of structure formation in $\Lambda$CDM cosmology, the dark matter subhalo population likely extends below the typical galactic mass scale ($\sim 10^{10} M_\odot$) down to $10^6 M_\odot$ or even smaller. Such small subhalos are optically invisible, but can have a pronounced effect on the lensing caustic strength while only modestly affecting the deflection and hence astrometric position of the lensed object \citep{Dai2018Abell370, Dai2020S1226millilens, Williams2024}. Indeed, \cite{JiDai2024} find through a combination of semi-analytic modeling and numerical ray-tracing that accounting for this population of subhalos relaxes the constraint on the source size on the visible side of the caustic derived in \cite{Welch2022b} to as large as $\sim3$~pc without contradicting observations. This light-weighted size is compatible with many of the massive star clusters seen both in the local universe \cite[for a review see][]{Krumholz2019} and at high redshift \citep[e.g.,][]{Adamo2024}, and has re-opened the discussion on the nature of Earendel and other similar highly magnified high-redshift sources.

Recently, \cite{Scofield2025} have also shown from a macroscopic analysis (i.e., not including the subhalos treatment of \cite{JiDai2024}) that Earendel's magnification may be as low as $\mu\sim50$. While \cite{JiDai2024} assume the strong-lensing models of \cite{Welch2022a} which predict magnifications in excess of 1000 and then subsequently evaluate how this changes in the presences of dark matter subhalos, \cite{Scofield2025} construct a new joint strong- and weak-lensing model which itself finds Earendel may only be modestly magnified. The approach of \cite{Scofield2025} was found to accurately predict magnifications for SN H0pe, the only multiply-imaged source in the galaxy cluster lensing regime with a known magnification in the literature \citep{Frye2024,Pascale2025}, and further draws the nature of Earendel into question.

Motivated by the conclusions of \cite{JiDai2024} and \cite{Scofield2025}, in this letter we consider whether Earendel
could instead be a compact star cluster rather than an individual star or star system, and simultaneously evaluate other stellar knots with in the Sunrise galaxy.
Sunrise is known to host multiple compact star clusters with exceptional stellar surface densities ($>10^4 M_\odot/{\rm pc}^2$) at both young ($<10$~Myr) and evolved ages \citep{Vanzella2023}. More generally, strongly-lensed galaxies have revealed an abundance of high density star clusters which are rare in our cosmic backyard \citep[e.g.,][]{Adamo2024, Whitaker2025,Messa2024,Fujimoto2024}. These clusters can probe vigorous star formation which may be more commonplace in the high redshift universe, and furthermore may be the progenitors of locally observed globular clusters whose origin and evolution are not yet fully understood \citep[e.g.,][]{Vanzella2017protoGCs}. Deep UV-Optical spectroscopic studies of these clusters are sparse, particularly for evolved cluster ages $>10$~Myr. At the redshift of Sunrise, deep publicly available NIRSpec spectroscopy covers the rest-UV through Optical of Earendel as well as multiple other stellar clumps within the galaxy which are more widely-accepted as star clusters. Hence Earendel and the Sunrise galaxy may be valuable assets for furthering our understanding of high redshift star clusters.

In section~\ref{sec:observations}, we describe the datasets used for this analysis, including archival {\it JWST} NIRSpec spectroscopy which has not been evaluated in prior literature on the Sunrise galaxy. Section~\ref{sec:meth} details the methods involved, including derivation of the spectroscopic redshift and the SED fitting to derive the physical properties. We discuss the results and the nature of Earendel as well as other stellar clumps in the Sunrise galaxy in section~\ref{sec:disc}. We evaluate whether the spectrum is consistent with the expectation of a star cluster, and assess how this system relates to the picture of star cluster formation and evolution. In section~\ref{sec:concl}, we provide concluding remarks on this analysis.

\section{Observations} \label{sec:observations}
\begin{figure*}
\centering
\includegraphics[scale=1.33]{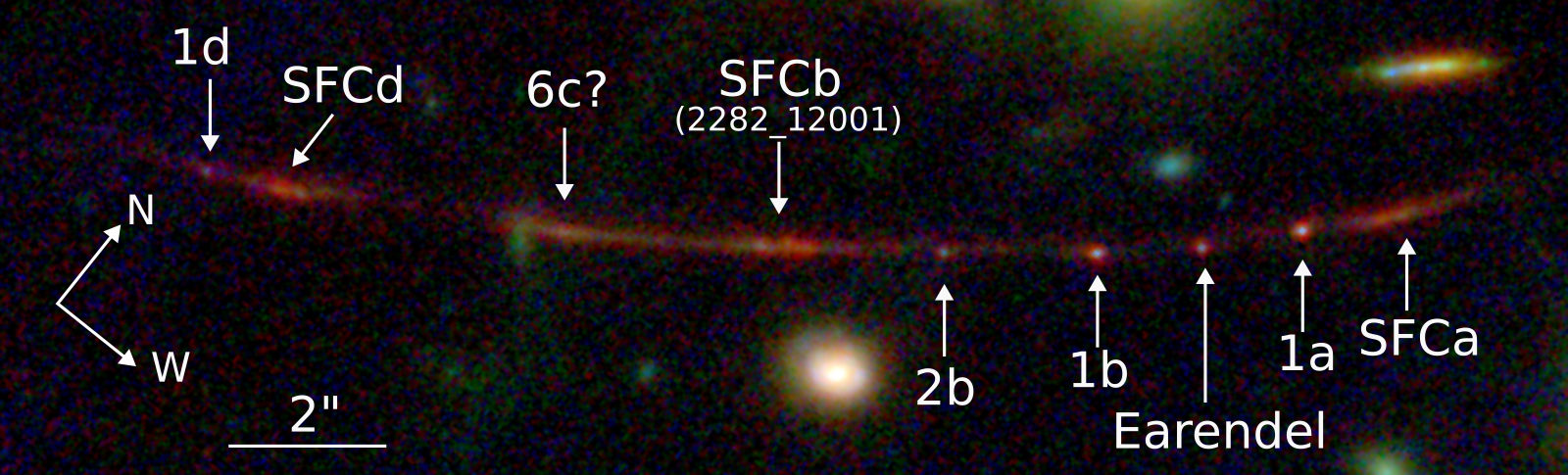}
\caption{{\it JWST} NIRCam F090W+F200W+F444W false color image of the Sunrise galaxy at $z\approx5.93$. The multiple images shown sweep out an angular extent of $\sim 16$ arcsec. At the redshift of Sunrise, the compass 2$\arcsec$ corresponds to $\sim 11.8$~kpc. Labeled are multiply-imaged clumps of interest within the galaxy: Earendel, image 1, image 6, and the so-called `star-forming complex', or SFC, consisting of images 3, 4, 5, and 6. We note the JWST NIRSpec MSA slit ID which covers image b of the SFC, \texttt{2282\_12001}, which is used for the spectroscopic redshift measurement.
}
\label{fig:earendel}
\end{figure*}

This work makes use  {\it JWST} datasets which are available on the MAST archive. {\it JWST} NIRCam imaging is not directly used in this analysis, however photometry results from \cite{Welch2022b} and \cite{Vanzella2023} which make use of images from Cycle 1 Go program 2282 (PI: Coe) are used. {\it JWST} NIRSpec PRISM spectra were queried from version 3 of the DAWN JWST Archive \citep[DJA, ][]{Heintz2024}, which make use of \texttt{msaexp} \citep{Brammer2022} for the reduction pipeline; further details can be found in \cite{Heintz2024} and \cite{deGraaff2025}.We highlight that the background subtraction is performed using a `master sky background' approach which fits to empty portions of the source slitlets and optimizes for signal to noise as described in Appendix A of \cite{deGraaff2025}. We make use of four spectra in the Sunrise for Earendel, image $1b$, image $1a$, and one of the star-forming regions along the Sunrise arc, which correspond to 2282\_10000, 2282\_10002, 2282\_10001, and 2282\_12001 respectively. The Sunrise galaxy and clumps of interest are shown in Fig.~\ref{fig:earendel}. The Earendel and $1b$ spectra are shown in Fig.~\ref{fig:spec}. We note that the singular emission-line spectrum, 2282\_1200, may contain a combination of images {\it 4, 5,} and {\it 6} which are denoted in \cite{Vanzella2023} as making up a so-called `star-forming complex' (SFC). This spectrum is only used for the spectroscopic redshift measurement, and as such the multi-component aspect of this spectrum is not accounted for in this work.

We adopt the nomencalture of \cite{Vanzella2023} for image system designations in this work (see our Fig.~\ref{fig:earendel} or their Fig.~2). However we note that the system of $d$ images of the SFC is somewhat misleading, as these images have an opposite parity to the system of $b$ images, indicating there must be an odd number of critical curve crossings through the Sunrise arc between the $b$ and $d$ systems. \cite{Scofield2025} similarly point out that the image system designations may be inaccurate. For simplicity, we retain the nomenclature used in previous works, however we caution that the image system designations are subject to change in future works.

\section{Methods} \label{sec:meth}

\subsection{Redshift Determination}\label{sec:redshift}
We determine the spectroscopic redshift of the Sunrise arc via nebular line emission detected in slit 12282\_12001, which contains SFCb on the arc (see Fig.~\ref{fig:earendel}) consisting of {\it 4b, 5b, and 6b} \citep[see nomenclature in ][]{Vanzella2023}. The spectrum shows clear detections of [\ion{O}{2}]$\lambda\lambda$3727,3729, [\ion{Ne}{3}]$\lambda$3869, 
H$\beta$, [\ion{O}{3}]$\lambda\lambda$4959,5007, and H$\alpha$ at SNR$\gtrsim$5.
The redshift is determined via simultaneous fitting of Gaussian line profiles to the H-Balmer and [\ion{O}{3}] features, yielding $z=5.926^{+0.013}_{-0.012}$. This is consistent with the Lyman-series break spectroscopically observed in the stellar continuum of other knots within the Sunrise (e.g., Earendel, 1a/b). Our spectroscopic measurement is lower than the photometric redshift $z_{\rm phot}= 6.2 \pm 0.1$ inferred by \cite{Welch2022b}, but consistent with the $z_{\rm phot} = 5.9^{+0.3}_{-0.2}$ determined by \cite{Vanzella2023}.

\subsection{SED Fitting}\label{sec:sed}
We perform Spectral Energy Distribution (SED) fitting to the continuum of Earendel and the lensed knot {\it 1b} to infer their physical parameters. We also perform fitting to the counterimage of $1b$, $1a$, whose spectrum is lower SNR but useful for testing the robustness of the fitting procedure as $1b$ and $1a$ should ideally yield comparable results. We choose to fit only to the NIRSpec spectrum rather than performing a joint photometric-spectroscopic fit. \cite{Welch2022b} performed detailed photometry of Earendel synthesizing 14 different measurements conducted by 10 separate people (a so-called ``wisdom of crowds'' approach), however we find the \cite{Welch2022b} measurements to disagree with the spectroscopic continuum at the $\gtrsim 20\%$ level in Earendel, and features an enhancement in the F115W filter flux which is not seen in the spectrum \citep[see Fig. 6 of ][]{Welch2022b}. Given the thoroughness of the \cite{Welch2022b} approach, we speculate that systematics may dominate the photometric error, or that the F115W filter may suffer from an artifact at the location of Earendel.

This photometric-spectroscopic tension is not seen in the neighboring knot, {\it 1b}, however as a caution we opt to fit only to the NIRSpec spectrum in all cases. While systematics from the spectroscopic calibration may affect the results, we find that the goodness-of-fit is reasonable in all spectroscopic fits, and furthermore that including photometry with the F115W filter removed generally agrees to within $1\sigma$. Stellar masses are measured by scaling to the F200W filter photometry from \cite{Welch2022b} and \cite{Vanzella2023}, with the exception of $1a$ which does not have existing photometry in the literature, and is only used to probe systematics in the SED fitting.

SED fitting is conducted by fitting a simple stellar population (SSP) to the observed NIRSpec spectra. The approach is modeled after the well-tested \texttt{Bagpipes} software \citep{Carnall2018,Carnall2019}, however our methodology is optimized for use of instantaneous burst star formation histories (SFH), which is likely appropriate for an individual star cluster \citep{Wofford2016}. We briefly describe the approach below, noting that the methodology follows \texttt{Bagpipes} unless specified otherwise. We also test an exponentially decaying SFH in Appendix~\ref{sfh}.

The stellar SED is constructed using three separate well-known SSP libraries to account for systematics involving the SSP used: 1) \texttt{BPASS} \citep{Stanway2016BPASS} including binary evolution and using an initial mass function (IMF) $\xi(m) \propto m^{-1.3}$ for $m<0.5\,M_\odot$ and $\xi(m) \propto m^{-2.35}$ for $m>0.5\,M_\odot$ in the initial stellar mass range $0.1 < m/M_\odot <100$. 2) The 2016 version of the BC03 \citep{Bruzual2003, Chevallard2016} models using a \cite{Kroupa2001} IMF and an upper-mass cutoff $M_{\rm cut} = 100 \,M_\odot$. And 3) \texttt{FSPS} \citep{Conroy2010} using the MIST \citep{Choi2016} isochrones with the MILES empirical stellar library and a \cite{Kroupa2001} IMF with an upper-mass cutoff  $M_{\rm cut} = 120\,M_\odot$.

For each SSP library, we construct a grid of ages and metallicites using values natively available from the library which we linearly interpolate across in $\log(t_{\rm age})$ and metallicity to yield the stellar SED. This is the primary difference between our approach and \texttt{Bagpipes}, which resamples the SSP age grid using a weighted summation approach \citep{Carnall2018}. In fitting, the population age is allowed to vary freely from 1~Myr to 1~Gyr (roughly the age of the universe at this redshift), and the metallicity is optimized in the range $Z/Z_\odot \in (0.0005,1)$ for \texttt{BPASS} and $Z/Z_\odot \in (0.005,1)$ for BC03 and FSPS \citep[where $Z_\odot=0.0142$; ][]{Asplund2021SolarAbund2020}.

We also include a nebular component with free ionization parameter $\log U\in (-4,-1)$ and free covering factor $x\in(0,1)$ which directly scales the nebular component flux. The nebular component is constructed using \texttt{CLOUDY} \citep{Ferland2017c17} with input incident radiation from the stellar SEDs which make up our grid. For each nebular realization, we fix the gas-phase metallicity to the metallicity of the input stellar population, assume rescaled solar chemical abundances (with the exception He and N which follow \cite{Dopita2000}), and use a fixed electron density $n_e=100{\rm ~cm}^{-2}$. Unlike \texttt{BAGPIPES}, we extend the nebular grid to include stellar populations out to $1$~Gyr, as stripped stars may provide a non-negligible ionizing budget through the first $\sim100$~Myr of the cluster lifetime \citep{Gotberg2019}.

We note that we identify candidate marginal detections of [\ion{O}{3}]$\lambda$5007 and H$\alpha$ in the Earendel spectrum which are measured to $2.5\sigma$ and $1.6\sigma$ significance respectively. Investigation of the 2D spectrum shows these features appear only marginally more extended than the underlying continuum, and we cannot rule out that they result from contamination from the arc background. As a caution we masked these features, but note that their inclusion was not found to impact the posterior of the fits.

We apply an ISM dust reddening law following \cite{Salim2018}, which includes a free reddening power law slope on which we force a prior $\delta \in [-2.0, 0.75]$. This has the freedom to reproduce both the \cite{Calzetti2001} ($\delta=0$) and SMC ($\delta\sim-0.45$) reddening curves, as well as significantly steeper curves $\delta<-1$ which may be better representative of low mass high redshift galaxies \citep{Salim2018}. We also include a free 2175\AA~ bump strength $B\in[0,3]$ where $B=3$ matches the Milky Way bump strength of \cite{Cardelli1989}. IGM attenuation and dust emission follow exactly the prescription of \texttt{Bagpipes} and we refer the reader to \cite{Carnall2018} for further details.

The system redshift was allowed to vary following a Gaussian prior with median and width $(\mu,\sigma)=(5.926,0.013)$ following the redshift determined in Sec.~\ref{sec:redshift}. The spectral resolution was set by the NIRSpec PRISM resolution curves provided by the {\it JWST} User documentation (JDox)\footnote{\href{https://jwst-docs.stsci.edu/files/97979440/97979447/1/1596073265467/jwst_nirspec_prism_disp.fits}{JWST NIRSpec PRISM Resolution Curve}} - as Earendel and image 1 are point-like sources, their position in the slit may affect the wavelength-dependent spectral resolution \citep{deGraaff2024}, however the resolution curve provided by JDox is likely a sufficient approximation for this analysis. We additionally allow for a velocity dispersion $\sigma \in (1, 100)$~km/s, encompassing the range of dispersions seen in even the most compact star clusters. Finally, we include an additional white noise scaling term $\alpha \in (1,10)$, which directly scales the NIRSpec spectrum uncertainties, and can help mitigate both underestimation of uncertainties and SED model systematics.

The resulting model consists of 10 free parameters for 407 observables with 397 degrees of freedom. Parameter inference is computed using \texttt{Nautilus} \citep{Lange2023} using 3000 live points. Robustness of the posterior is ensured following recommendations from \cite{Lange2023}, specifically that: 1) increasing the number of live points does not change the result, 2) points from the exploration phase are discarded, and 3) the shell-bound occupation fraction is unity beyond the earliest iterations.

\begin{figure*}
\centering
\includegraphics[scale=0.55,trim={0 2cm 0 0},clip]{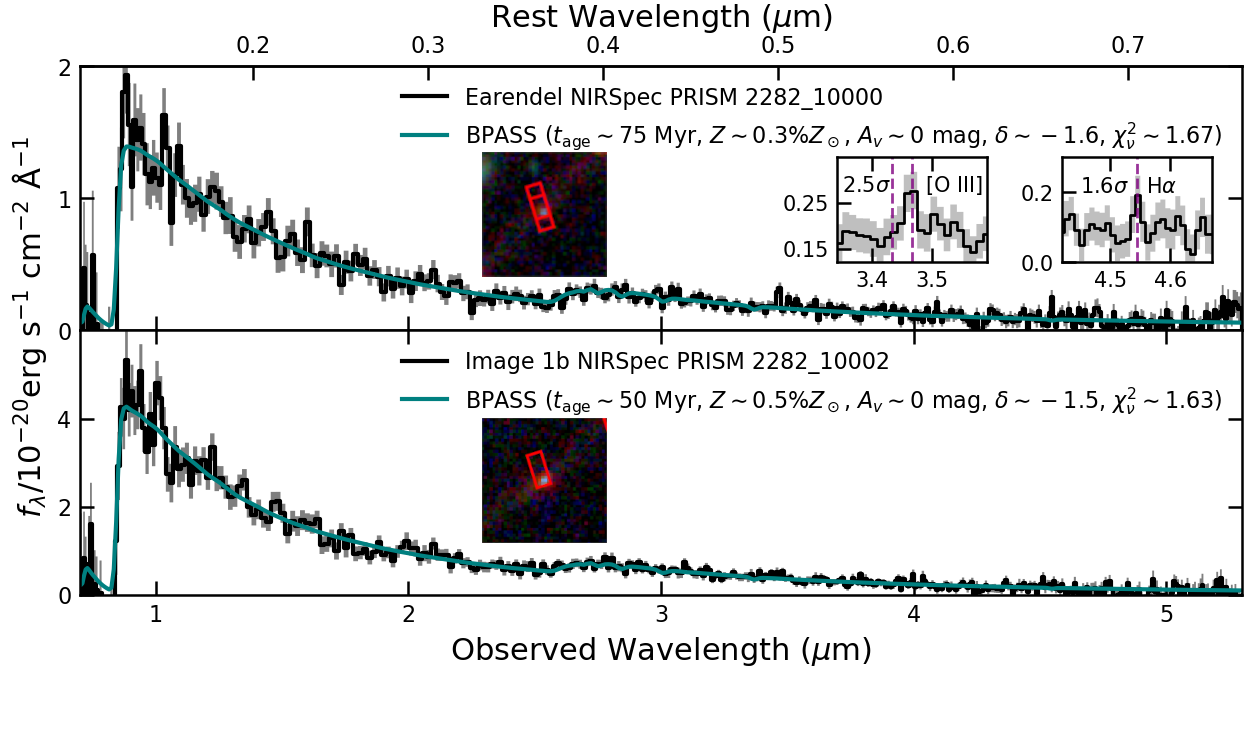}
\caption{NIRSpec PRISM spectra of Earendel and $1b$ with the best fit \texttt{BPASS} model from SED fitting overplotted. We omit the BC03 and \texttt{FSPS} best fit SEDs for clarity, as they appear similar to the result from \texttt{BPASS}. Spectra are taken from version 3 of the DJA \citep{Heintz2024}. Both Earendel and $1b$ have similar spectra indicative of stellar continuum from an evolved simple stellar population. The fit is influenced by the shape of the continuum, the lack of significant absorption 
features, as well as the clear Balmer break seen in both spectra. SED fitting measures similar parameters for these two objects, which are labeled approximately. Left insets show stamps from a {\it JWST} false color image with the NIRSpec MSA slits overplotted. Right insets show smoothed zoom-ins of the Earendel spectrum where we identify marginal detections of nebular line emission [\ion{O}{3}]$\lambda$5007 ([\ion{O}{3}]$\lambda$4959 non-detection also marked) and H$\alpha$ which are measured at $2.5\sigma$ and $1.6\sigma$ significance respectively.
}
\label{fig:spec}
\end{figure*}

We perform two primary tests to address underestimation of uncertainties beyond the inclusion of the white noise scaling term. We first test if outlier features, such as those at $\sim 10500{\rm \AA}$  observer frame in the Earendel spectrum (see Fig.~\ref{fig:spec}), skew the fit by masking them, which we did not find to meaningfully impact the posteriors. We also test in the limiting case where the intrinsic spectrum is mostly smooth, by applying a boxcar smoothing function from the spectrum redward of the Lyman-series limit. We subtract the smoothed spectrum from the original, which was found to approximately center at zero, and then calculate a running standard deviation across the spectrum. We find this running standard deviation to be consistent with the uncertainties applied in fitting after accounting for the best fit white noise scaling. Furthermore, recomputing the fit assigning the running standard deviation as the error was found to produce similar results to the assigned errors.

\section{Results and Discussion} \label{sec:disc}

\subsection{Metal Poor Globular Cluster Progenitors} \label{sec:gc}
We find the Earendel spectrum to be highly consistent with that of a simple stellar population across all three SSP libraries. Each library produces a comparable best-fit, in each case requiring a white noise scaling $\sim 1.3$ to reach a reduced chi-squared of unity, translating to a typical $\chi^2_\nu\sim 1.7$ without the noise scaling term. We similarly a nearly equivalent goodness-of-fit for image $1b$, which is commonly accepted to be a star cluster. We reassuringly find that repeating the same fitting procedure for the $1a$ spectrum yields comparable results to $1b$, as seen in Fig.~\ref{fig:corner} and Table~\ref{tab:sed}. The results of \texttt{BPASS} and BC03 strongly agree to within $1\sigma$, while the \texttt{FSPS} fits show a weaker $\sim 2\sigma$ level agreement but remain qualitatively similar.

The best fit spectrum from the \texttt{BPASS} templates is shown in Fig.~\ref{fig:spec}. All three SSP libraries produce qualitatively similar posteriors for both Earendel and $1b$, however some physical parameters disagree to $\gtrsim 2\sigma$; the inferred parameters are given in Table~\ref{tab:sed} and the \texttt{BPASS} posteriors are plotted in Fig.~\ref{fig:corner}. For both Earendel and $1b$, we infer an evolved ($>30$~Myr), metal-poor ($Z/Z_\odot <10\%$) star cluster with a magnified mass $\mu M \sim 10^9\,M_\odot$ and a minimal ISM dust reddening $A_{\rm V}<0.1$~mag.  We do not find the velocity dispersion to be well-constrained, nor does it significantly impact the posteriors, reflecting both the low spectral resolution of PRISM and the lack of significant absorption features.

\subsubsection{Stellar Surface Density}

The Sunrise galaxy is well-know to host some of the highest stellar surface density clusters observed, and indeed \cite{Vanzella2023} determine from photometrically derived stellar masses and astrometrically derived sizes the stellar surface density of $1b$ to be $\Sigma_\star \approx 3\times10^5 (\frac{\mu}{70}) {\rm M}_\odot {\rm ~pc}^{-2}$. We find our updated spectroscopically derived stellar masses are higher than those determined by \cite{Vanzella2023}, a consequence of generally older derived ages. Assuming the magnification and size estimates of \cite{Vanzella2023} and our BPASS results (the intermediate of the three SSP mass estimates) of ${M}_\star/{\rm M}_\odot \approx 2.5\times10^7 (\frac{\mu}{70})$, we estimate the $1b$ stellar surface density to be $\Sigma_{\star,1b} \approx 1.3 \times 10^{6}  (\frac{\mu}{70})  {\rm M}_\odot {\rm ~pc}^{-2}$.

The stellar surface density measurement for Earendel is less straightforward, as the magnification and size measurements are not well-understood. \cite{Welch2022b} find the apparent surface brightness of Earendel to be approximately consistent with the PSF in {\it JWST} NIRCam F090W imaging, placing the apparent source size below that of a native JWST pixel (0.031$\arcsec \sim$ 200~pc at z=5.926).

In the simulations of \cite{JiDai2024}, the light of Earendel is dominated by a single lensed image, and the intrinsic source size may be as large as 1-3~pc while still appearing below this $0.031\arcsec$ limit. Additionally, new strong lensing models by \cite{Scofield2025} find macroscopically (e.g., without subhalos) that Earendel's magnification may be as low as $\mu\sim40-70$, further complicating the choice of magnification.

We consider a range of fiducial tangential magnifications $\mu_t\sim50-500$ with an assumed radial magnification $\mu_r\sim2$ , producing FUV size upper limits $r_{\rm FUV} \approx 0.4-4\,{\rm pc}$ and star cluster of masses $M_*/M_\odot \approx 10^{6-7}$ assuming the \texttt{BPASS} posterior median. The resulting stellar surface density upper limit is hence $\Sigma_{\star,{\rm e}} \approx 3\times10^{5-6}  {\rm M}_\odot {\rm ~pc}^{-2}$.

Both Earendel and $1b$ produce incredible stellar surface densities, which for some magnifications exceed the observed $\Sigma_{\star,{\rm max}}\approx 3\times10^5  {\rm M}_\odot {\rm ~pc}^{-2}$ ceiling seen in local universe clusters \citep{Hopkins2010MaximumSurfaceDensity,Grudic2019}. Such densities may indicate the star-formation in these clusters is feedback free, allowing for higher star formation efficiency and hence higher stellar densities \citep{Dekel2023}. However \cite{Williams2025} determine that in such a scenario, clusters of stellar surface density $\log(\Sigma_\star)>10^5$ and ${\rm M}_\star > 10^6 {\rm M}_\odot$ are typically bound yet non-virialized; such high densities may not be maintained across the cluster evolution to the present day. Highly efficient star-formation may also be conducive for the formation of populations with top-heavy IMFs \citep[e.g.,][]{Menon2024, Lake2025}, however we did not find that implementing SSPs with a top-heavy IMF ($\xi(m)\propto m^{-2.00}$ for $m>0.5M_\odot$) meaningfully changed the stellar mass estimates. Finally, we note that we derive these stellar surface densities using rest-FUV sizes, which due to mass segregation \citep[e.g.,][]{Hosek2015,Mestric2023VMS} may be smaller than the optical sizes typically quoted for local universe clusters or those quoted from simulations.
\vspace{5mm}
\subsubsection{Age and Metallicity}
\begin{figure*}[t]
    \centering
    \includegraphics[scale=0.29]{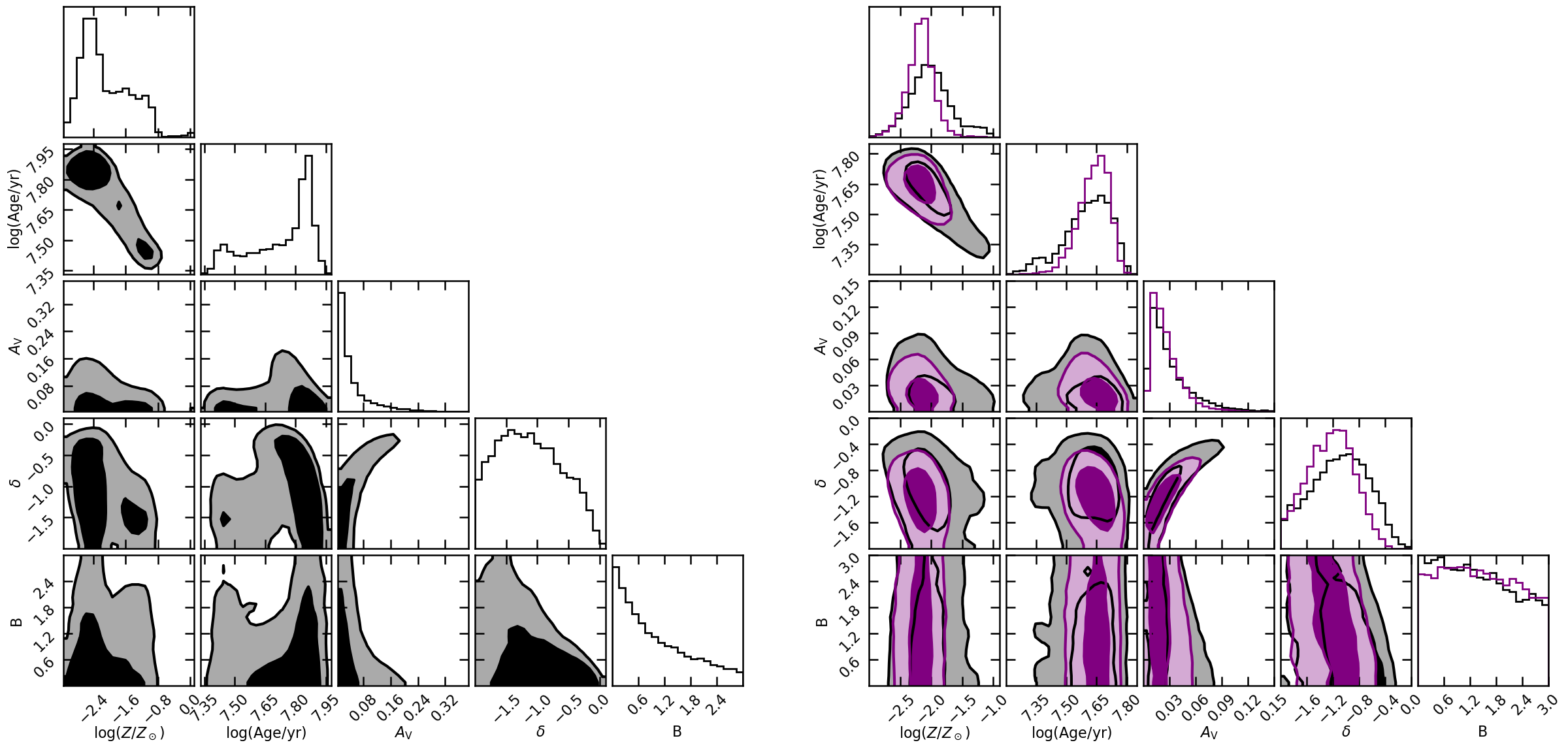}
    \caption{Posterior distributions for the Earendel (left), $1b$ (right purple), and $1a$ (right black) physical parameters from fitting with the \texttt{BPASS} SSP library, which generally yielded the largest age and metallicity uncertainties. We omit the ionization parameter, covering factor and velocity dispersion as they go unconstrained. We also omit the white noise scaling, which is $\lesssim1.3$ in all cases. 2D contours enclose 50\% and 90\% of the posterior samples. Refer to Table~\ref{tab:sed} for the 68\% C.I.'s of the parameters.
    }
    \label{fig:corner}
\end{figure*}

All three SSPs show general agreement at the $\sim 1\sigma$ level in their metallicity predictions for both Earendel and $1b$, and consistently predict the metallicity to be in the metal-poor regime. Choosing whichever SSP posterior predicts allows the highest metallicities (generally \texttt{BPASS}), the inferred posteriors rule out $Z/Z_\odot > 10\%$ to $\sim$95\% confidence in Earendel and $\sim$99\% confidence for $1b$. Similarly $Z/Z_\odot > 5\%$ is ruled out to $\sim$80\% and  $\sim$ 99\% confidence for Earendel and $1b$ respectively. We  point out that the uncertainties quoted for the \texttt{BPASS} fit of Earendel are somewhat misleading. Fig.~\ref{fig:corner} reveals the age-metallicity posterior is closer to double-peaked rather than classically broad, and metal-poor solutions are strongly favored.

We found these metallicity measurements were influenced primarily from the shape of the spectroscopic continuum (e.g, the UV and optical slopes, the observed Balmer break), as the depth of absorption features was not found to approach the flux uncertainties until $Z\sim 0.5\,Z_\odot$. While we are cautious to assign an exact metallicity to either Earendel or $1b$ given the potential systematics, the synthesis of the three SSP fits confidently places these clusters in the metal-poor regime ($\lesssim 10\% Z_\odot$). Further tests assessing systematics in the metallicity measurement are detailed in Appendix~\ref{Testmet}.

The ages are in weaker agreement with one another compared to the metallicity. For Earendel the SSPs predict a wide range of ages $30-110$~Myr, and the uncertainties are of order 10's of Myr. In the case of $1b$, the statistical precision is much higher (noting that uncertainties are scaled up $\sim30$\% via a white noise term), with BC03, \texttt{BPASS}, and \texttt{FSPS} predicting $t_{\rm age}\sim 30, 45$ and $160$~Myr respectively to within a $\sim$few~Myr precision. This may be reflective of the overall slow evolution of the stellar continuum beyond $10$~Myr, and perhaps somewhat poorly understood ingredients in the SSP such as the treatment of binaries (though we note that repeating fitting with the \texttt{BPASS} single star models was not found to impact the solution) or the evolutionary tracks of red supergiants \citep[e.g., ][]{Vazquez2005}. Indeed visual inspection shows that crucial spectral features for this work, such as the Balmer break strength, are not in always in strong agreement across the 10-100~Myr age range in these SSPs.

Observations out to redder wavelengths may help narrow both the age and metallicity estimates. Specifically, observations of the broad photospheric $1.6\mu$m H$^-$ opacity feature could be leveraged to age-date star clusters in the $>10$~Myr range due to red supergiants at early times \citep{Eldridge2020} and later red giants and AGB stars \citep[e.g.,][]{Sawicki2002}. This feature is most apparent at low metallicities \citep{Yang2021}, and is broad enough to be identifiable photometrically with {\it JWST} MIRI imaging.

\begin{figure}[h]
\includegraphics[scale=0.375]{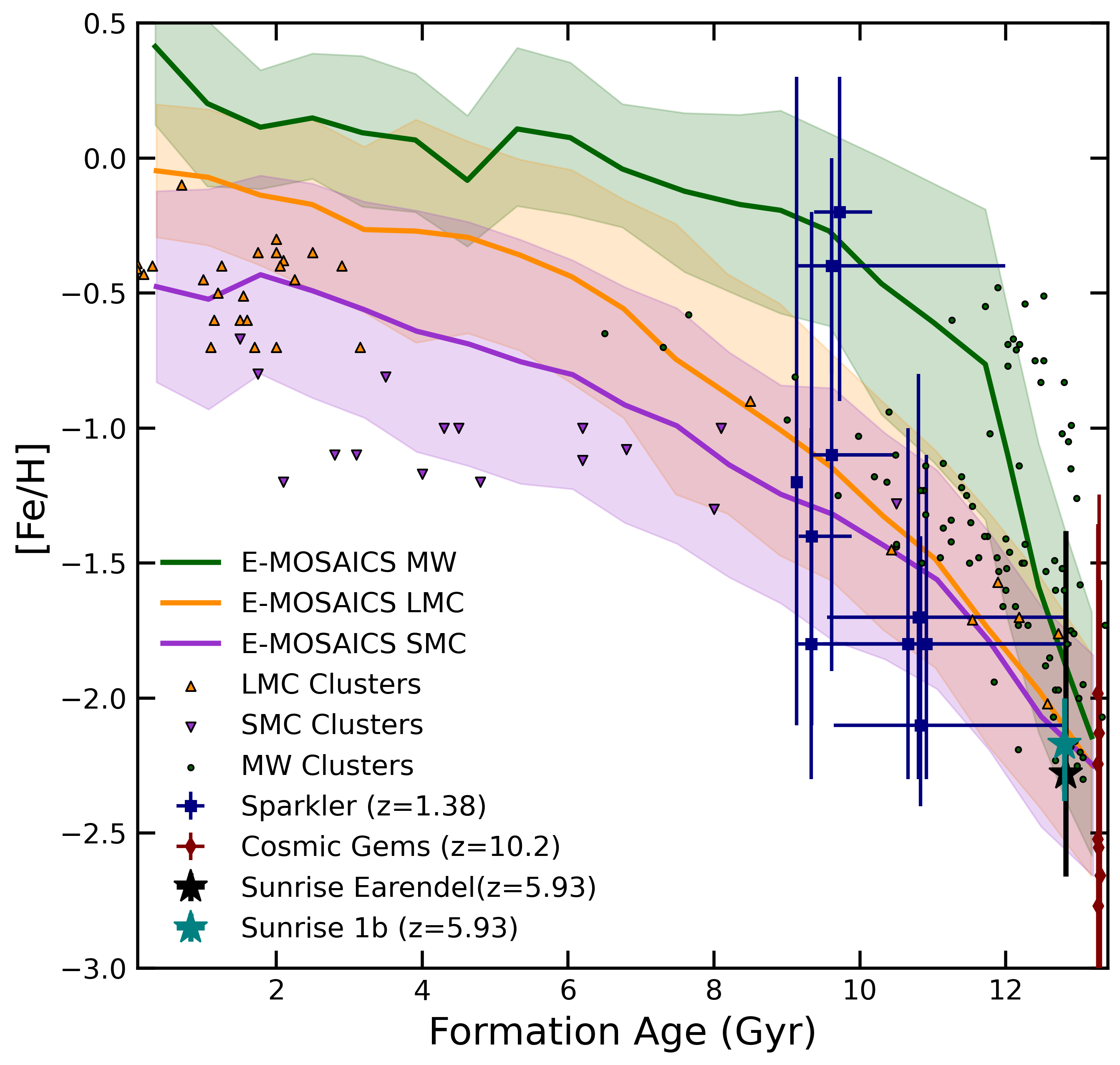}
\caption{Massive star cluster age-metallicity relation predicted for MW, LMC, and SMC mass host galaxies predicted by the E-MOSAICS cosmological simulation \citep{Kruijssen19, Horta2021} compared to measurements from LMC and SMC clusters \citep[see Table A1 of][and references therein]{Horta2021}, MW clusters \citep[see Table A1 of][and references therein]{Kruijssen19}, from the Sparkler \citep[$z=1.38$;][]{Adamo2023}, the Cosmic Gems \citep[$z\approx10.2$;][]{Adamo2024}, and Earendel and $1b$ (\texttt{BPASS} results which have the largest uncertainties) from this work. We note that the younger ages of some clusters within both the Sparkler and the Cosmic Gems implies the best-fit metallicity is dictated in part by enhancement from nebular emission, however we still 
approximate $\log(Z/Z_\odot) \approx {\rm [Fe/H]}$.
All three sets of clusters appear broadly consistent with the predictions of SMC/LMC mass host galaxies, with the Sunrise also showing consistency with a MW mass host. This figure is adapted from Fig.~4 of both \cite{Adamo2023} and \cite{Horta2021}.
}
\label{fig:comp}
\end{figure}

The cosmic formation epoch and metallicities of Earendel and $1b$ are consistent with the age-metallicity sequence observed from globular clusters in local galaxies \citep[e.g., ][also see Fig.~\ref{fig:comp}]{Cezario2013, Usher2019}. Furthermore, these findings appear consistent with the ongoing picture of star cluster evolution indicated by high redshift proto-GCs. Earendel and $1b$ may be evolved cousins of the clusters seen in the $z\sim 10$ Cosmic Gems~\citep{Adamo2024} and the $z\sim 8.3$ Firefly Sparkle \citep{Mowla2024}, while perhaps being the precursors to older proto-GCs seen at more intermediate redshifts such as the $z\sim1.4$ Sparkler \citep{Mowla2022,Adamo2023} or the $z\sim2.5$ Relic \citep{Whitaker2025}.

In Fig.~\ref{fig:comp}, we compare the cosmic formation ages and metallicities of various high redshift clusters to predictions of the cluster age-metallicity relation (AMR)  for MW, LMC, and SMC-like galaxies presented by \cite{Horta2021} from the E-MOSAICS simulation \citep{Pfeffer2018, Kruijssen19}, where the MW relation is derived from zoom-in simulations in \cite{Kruijssen19} and the LMC/SMC relations are from \cite{Horta2021}. While Earendel and $1b$ show best agreement with SMC/LMC-like halos, the Sunrise galaxy may actually be closer to a MW-like halo mass. \cite{Vanzella2023} derive a magnification-corrected total stellar mass $M_\star \sim 10^{8-9}M_\odot$, which is as much as an order of magnitude higher than the LMC had at this cosmic age \citep{Weisz2013}. Given the low metallicities inferred for these clusters, it is not immediately clear if they more closely resemble the in-situ or ex-situ branch of GCs seen in the Milky Way \citep{Forbes2010}. Should Earendel's true metallicity correspond to the lower age, higher metallicity solution seen in Fig.~\ref{fig:corner}, this would place it more squarely in the in-situ branch, reflecting the rapid build of up metals achievable in a Milky Way mass halo \citep[e.g.,][]{Kobayashi2020}

\begin{deluxetable*}{ccccccccc}
\tabletypesize{\footnotesize}
\tablecaption{SED Parameters}
\label{tab:sed}
\tablecolumns{9}
\tablehead{
\colhead{\bf Image} &  \colhead{\bf SSP} &  \colhead{\bf $\log(t_{\rm age}/{\rm yr})$} &  \colhead{\bf $\log(\mu M/M_\odot)$}  & \colhead{\bf $\log(Z/Z_\odot)$} & \colhead{$A_{\rm V}$~(mag)} & \colhead{$\delta$} & \colhead{$B$} & \colhead{\bf $\chi^2_\nu$}}
\startdata
Earendel & BC03 & $7.50^{+0.04}_{-0.04}$ & $8.87^{+0.03}_{-0.04}$ & $-1.67^{+0.26}_{-0.21}$ & $0.03^{+0.03}_{-0.01}$ & $-1.44^{+0.44}_{-0.35}$ & $0.98^{+1.10}_{-0.71}$ & $1.71$ \\
 & \texttt{BPASS} & $7.79^{+0.08}_{-0.27}$ & $9.06^{+0.07}_{-0.23}$ & $-2.28^{+0.90}_{-0.38}$ & $0.03^{+0.06}_{-0.02}$ & $-1.15^{+0.61}_{-0.52}$ & $0.77^{+1.17}_{-0.59}$ & $1.67$ \\
  & \texttt{FSPS} & $8.05^{+0.14}_{-0.27}$ & $9.24^{+0.18}_{-0.23}$ & $-1.67^{+0.48}_{-0.40}$ & $0.05^{+0.14}_{-0.04}$ & $-0.09^{+0.49}_{-0.93}$ & $0.73^{+1.19}_{-0.56}$ & $1.68$ \\
\hline
Im 1b & BC03 & $7.47^{+0.01}_{-0.02}$ & $9.21^{+0.02}_{-0.02}$ & $-2.12^{+0.05}_{-0.03}$ & $0.03^{+0.02}_{-0.01}$ & $-1.25^{+0.29}_{-0.30}$ & $1.57^{+0.94}_{-0.97}$ & $1.65$ \\
 & \texttt{BPASS} & $7.64^{+0.06}_{-0.08}$ & $9.33^{+0.04}_{-0.06}$ & $-2.17^{+0.18}_{-0.21}$ & $0.02^{+0.02}_{-0.01}$ & $-1.22^{+0.34}_{-0.39}$ & $1.40^{+1.02}_{-0.93}$ & $1.63$ \\
 & \texttt{FSPS} & $8.21^{+0.03}_{-0.05}$ & $9.73^{+0.05}_{-0.05}$ & $-2.41^{+0.12}_{-0.06}$ & $0.02^{+0.05}_{-0.02}$ & $0.30^{+0.33}_{-0.61}$ & $1.42^{+1.03}_{-0.97}$ & $1.65$ \\
\hline
Im 1a & BC03 & $7.47^{+0.02}_{-0.02}$ & $^{a}9.07^{+0.02}_{-0.01}$ & $-2.10^{+0.07}_{-0.04}$ & $0.03^{+0.02}_{-0.01}$ & $-1.29^{+0.33}_{-0.35}$ & $1.49^{+0.97}_{-0.96}$ & $1.77$ \\
 & \texttt{BPASS} & $7.62^{+0.10}_{-0.13}$ & $^{a}9.17^{+0.08}_{-0.11}$ & $-2.03^{+0.33}_{-0.31}$ & $0.02^{+0.03}_{-0.01}$ & $-1.08^{+0.43}_{-0.48}$ & $1.32^{+1.08}_{-0.92}$ & $1.72$ \\
 & \texttt{FSPS} & $8.01^{+0.11}_{-0.13}$ & $^{a}9.42^{+0.14}_{-0.11}$ & $-1.81^{+0.33}_{-0.40}$ & $0.04^{+0.07}_{-0.03}$ & $0.30^{+0.32}_{-0.51}$ & $1.26^{+1.13}_{-0.90}$ & $1.77$ \\
\enddata
\tablecomments{Physical parameters for Earendel, image $1b$, and image $1a$ of the Sunrise arc using three separate SSP libraries: the 2016 version of the \cite{Bruzual2003}, BC03, models \citep{Chevallard2016}, \texttt{BPASS} \citep{Stanway2016BPASS}, and \texttt{FSPS} \citep{Conroy2010}. We list the object, the SSP library, the burst age, the magnified stellar mass ($\mu$ is the flux magnification factor due to lensing), the stellar metallicity, visual band extinction, reddening law slope following \cite{Salim2018}, reddening law 2175~\AA bump strength following \cite{Salim2018}, and the reduced chi-squared statistic for 397 degrees of freedom without the $\sim 30\%$ white noise scaling applied. While the nebular ionization parameter and covering factor are also optimized for in this analysis, they are omitted from the table for clarity as they go unconstrained within their uniform priors listed in Sec.~\ref{sec:meth}.}

\tablenotetext{a}{The $1a$ magnified mass is derived solely from the spectrum and not via rescaling to the measured F200W flux, as is the case for Earendel and $1b$. The $1a$ spectrum suffers from slit loss, and the mass is likely underestimated by a factor of 2-3.}
\end{deluxetable*}

\subsection{Dust Reddening and Nebular Parameters} \label{sec:reddening}
Across the three SSP inferences of Earendel and $1b$, we find minimal dust reddening $A_{\rm V} \lesssim 0.1$~mag. The reddening law slope $\delta$ is not well-constrained; the \texttt{BPASS} and BC03 fits loosely preferring slopes steeper than SMC-like (corresponding to $\delta=0.45$) however \texttt{FSPS} generally prefers slopes Calzetti-like ($\delta=0$) and shallower. The former results may be more consistent with the expectation that high redshift star-forming galaxies should have steeper reddening laws, for example the relations of \cite{Salim2018} find a slope of $\delta\sim-0.8$ for high redshift analogs of stellar mass $M_\star/M_\odot \sim 10^8$ similar to the Sunrise \citep{Vanzella2023}. By contrast, \cite{Sanders2024b} found the nebular attenuation curve of high star-forming galaxy at $z=4.41$ to have an FUV-Optical slope similar to \cite{Cardelli1989}, more closely matching the \texttt{FSPS} result. We also do not see a preference for a strong 2175~\AA~bump, but it is weakly disfavored for all fits to Earendel. Higher SNR is necessary to more confidently rule out the presence of a bump.

None of the fits provide any significant constraints on the ionization parameter nor the covering factor, reflecting the preference for evolved cluster ages which have lost the bulk of their ionizing budget. As noted in Sec.~\ref{sec:meth}, we identify candidate detections of [\ion{O}{3}]$\lambda5007$ and H$\alpha$ in the nebular spectrum. Given their low statistical significance, we refrain from making any definitive statements about their nature - the 2D spectrum does not clearly confirm nor deny them to be associated with the underlying arc, and they indeed may merely be a statistical fluke altogether.

The lack of prominent dust reddening and nebular emission is perhaps consistent with the expectation evolved star clusters are nearly devoid of gas \cite{Lada2003}, as feedback from massive stars is typically expected to expel nearly all of the cluster gas after the first $\sim$10-20~Myr of the cluster lifetime \citep{Calura2015WindFeedback}. While SNe and Wolf-Rayet stars are expected to provide the bulk of the cluster's dust production within first $\sim10$~Myr \citep{Schneider2024}, it is possible their dusty ejecta is not efficiently retained within the cluster, or that their overall covering is fairly low. 

We note more generally that the use of the \cite{Salim2018} reddening law is impactful for capturing uncertainties on the dust and age constraints in this work. Prior work by \cite{Nakane2025} has pointed out that, for a fixed \cite{Calzetti2001} reddening law, the age, dust, and metallicity affect the continuum in distinct ways, reducing the presence of a degeneracy between these three parameters. However the dust reddening law of high redshift galaxies is not necessarily Calzetti-like nor is it standardized \citep[e.g.,][]{Fisher2025}, and assuming a fixed slope could induce a bias in the inference or underestimate the uncertainties. Indeed, we find in our fits the flexible \cite{Salim2018} reddening law has a modest, but statistically relevant impact on the metallicity compared to implementing a fixed reddening law, even in the regime of $A_V<0.1$. This is most apparent in the \texttt{BPASS} fit to Earendel shown in Fig.~\ref{fig:corner}, where a secondary low age, high metallicity solution appears only for steep slopes $\delta<-1$. Given the broad posterior inferred on the reddening law slope $\delta$ in all of our fits, we find implementing the \cite{Calzetti2001} or SMC \citep{Gordon2003} reddening laws generally still agree with our \cite{Salim2018} fits to within $1\sigma$, however in a higher SNR regime this is likely to no longer be the case. As the dust reddening slope optimized for in this work, we acknowledge that it does not fully eliminate systematic bias related to the slope, however we find it is reduced compared to assuming the fixed slope apriori.

\subsection{Is Earendel a Star or Star Cluster?} \label{sec:star}

Prior work by \cite{Welch2022b} found the {\it JWST} photometry of Earendel was inconsistent with single star models from the PoWR and TLUSTY models. Rather the photometric data is better fit by a binary system, consisting of two massive ($m_{\rm ZAMS}\gtrsim 20\,M_\odot$) stars: an evolved supergiant with $T_{\rm eff,1}\sim 9000$~K, and a main sequence O star of $T_{\rm eff, 2}\sim 34000$~K. Assuming the largest magnification inferred across all existing lens models for the system, $\mu \sim 17000$, the total bolometric luminosity for Earendel must reach $L_{\rm bol}/L_\odot \sim 10^6$. \cite{Welch2022b} find the hotter star has a bolometric luminosity $L_{\rm bol}/L_\odot \sim 10^{5.8}$, $\sim 0.5$~dex greater than the cooler star. Such a system may be somewhat unusual compared to binary systems typically observed in local galaxies, as the hotter, brighter O star would be expected to leave the main sequence prior to its lower mass companion. Furthermore, the stellar parameters of the cooler star are indicative of passing through the short-lived ($\lesssim 10^4$~yr)  yellow supergiant phase \citep[e.g.,][]{Drout2009}, however the longer-lived ($\sim$~Myr) phase of a cool blue supergiant may also match the parameters.

This does not definitively rule out the binary scenario (see \cite{Nabizadeh2024arXiv240612607N} for possible solutions). While it is possible a single or binary stellar SED fit to the NIRSpec spectrum may provide results more consistent with local observations, it more generally may be difficult to confirm a binary or single-star scenario based on the spectrum alone. The hotter star in the \cite{Welch2022b} fit would likely present strong stellar wind features in emission such as the \ion{C}{4}$\lambda$1550 or \ion{N}{4}$\lambda$1240 P-Cygni features given an SMC-like metallicity (undetectable at the PRISM resolution), which would be highly inconsistent with an evolved star cluster. However if the system is instead metal poor, the wind mass loss rate and hence strength of these wind features would be too weak to be detectable in even deep high resolution spectra \citep[e.g.,][]{Leitherer2014, Telford2024}. By contrast, the cooler star would present stellar absorption features, such as the Balmer series or \ion{Ca}{2} H\&K, however these would also be consistent with an evolved star cluster which consists of many cool stars without strong winds. Ultimately, detection of microlensing-induced photometric variability may be the most promising `smoking gun' indicator \citep{Welch2022a}, however multi-epoch imaging has not yet confirmed such variability \citep{Welch2022b}.

A star cluster appears to be a natural conclusion from the spectroscopic data, and provides an exceptional fit across three different SSP libraries. The physical parameters estimated for the star cluster are consistent with the expectations of a globular cluster progenitor at $z\sim6$, and are in agreement with the results of simulation work on star cluster formation and evolution. The cluster age is also in alignment with the non-detection of statistically significant variability across the 2 year observer-frame baseline \citep{Welch2022b}, as such an evolved cluster lacks the massive stars necessary for microlensing to induce large fluctuations \citep{Dai2020S1226millilens}.

A third, but less well-defined scenario could be that Earendel consists of a ``small group'' of dozens of stars, requiring a large ($\mu>1000$) but not extreme ($<10^4$) magnification. Such a group likely cannot be sampled from a standard IMF, and must instead be a young system consisting of luminous O and B supergiants in order to reach the necessary bolometric luminosity $L_{\rm bol} \sim 10^7 L_\odot$ for $\mu \sim \mathcal{O}(10^3)$. Such a system in isolation is not seen in the local universe, and furthermore it is unclear whether it would match the observed spectrum without significantly fine-tuning the stellar members. However given the overall small number of stars, microlensing induced variability should be measurable through precision {\it JWST} photometry \citep{Dai2021Sunburst}, and stellar wind features may be detectable in medium resolution {\it JWST} spectroscopy given the system is not metal poor.

\section{Conclusion} \label{sec:concl}
In this letter, we evaluated two stellar clumps within the $z\approx5.93$ lensed Sunrise galaxy: Earendel, which we hypothesize is a star cluster, and another star cluster within the galaxy, $1b$. The quality of rest-UV through optical {\it JWST} NIRSpec spectra of these objects enabled detailed physical parameter estimation of these star clusters. Through SED fitting with three well-tested simple stellar population libraries, BC03, \texttt{BPASS}, and \texttt{FSPS}, we inferred their ages, metallicities, masses, and dust extinctions. We found both have intermediate ages, $t_{\rm age}\sim30$--$150$~Myr, and inferred their stellar metallicities are $Z \lesssim10\%\,Z_\odot$, as low as $Z\lesssim 1\%\,Z_\odot$ depending on the SSP model.

The formation ages and metallicities of these clusters are in relative agreement with the age-metallicity relation of globular clusters both measured in local galaxies and recovered in simulations. The results are in best agreement with the expectations of SMC/LMC halo mass host galaxies with more marginal consistency with a MW mass halo, which given the stellar mass of Sunrise ($M_\star \sim 10^{8-9}M_\odot$) is likely more appropriate. We have concluded that, when taken together with other high-redshift star cluster measurements, such as the Sparkler or the Cosmic Gems, these clusters appear to corroborate the picture of globular cluster formation and evolution illustrated by local GCs and simulations.

This study represents a first for spectroscopic fitting of evolved star clusters covering from rest-frame UV to optical wavelengths at these redshifts. Past studies have typically relied on nebular emission lines, which is only practical for very young ages, or solely photometry for ascertaining high redshift star cluster characteristics. While we do not find complete agreement between the three SSP libraries used, we find the low metallicity and intermediate ages inferred to be consistent results. Given that these clusters are extremely faint $m_{\rm F200W}\sim 27$~AB mag, the quality of these measurements showcases the power of {\it JWST}, and indicates that spectroscopic analysis of evolved star clusters in other high redshift lensed galaxies are viable.

\section*{acknowledgments}
The JWST data presented in this article were obtained from the Mikulski Archive for Space Telescopes (MAST) at the Space Telescope Science Institute. The specific observations analyzed can be accessed via \dataset[doi: 10.17909/gts7-8629]{https://doi.org/10.17909/gts7-8629}. First and foremost we would like to thank the anonymous referee for their helpful comments. We would like to thank Maude Gull, Alessandro Savino, Dan Stark, Liangyuan Ji, Chema Diego, Dan Coe, Brian Welch, Jan Eldridge and Masamune Oguri
for the very valuable discussions and comments. We thank Danny Horta for providing the E-MOSAICS results. We thank Dan Coe and the team behind {\it JWST} Cycle 1 program 2282 for leading these observations and waiving the proprietary period. Finally, we thank the Dawn JWST Archive (DJA), an initiative of the Cosmic Dawn Center (DAWN) which is funded by the Danish National Research Foundation under grant DNRF140. MP acknowledges acknowledges the support of System76 for providing computer equipment. L.D. acknowledges research grant support from the Alfred P. Sloan Foundation (Award Number FG-2021-16495), from the Frank and Karen Dabby STEM Fund in the Society of Hellman Fellows, and from the Office of Science, Office of High Energy Physics of the U.S. Department of Energy under Award Number DE-SC-0025293.
\bibliography{earendel}{}
\bibliographystyle{aasjournal}



\begin{appendix}
\twocolumngrid
\section{Tests of Star-formation History} \label{sfh}
For the SED fitting in this work, we assume the star formation history of these star clusters is representative of an instantaneous burst, which is appropriate for such dense star clusters which likely form their stars over only a few free-fall times of the parent giant molecular cloud \citep[$t_{\rm ff}\sim$~3 Myr assuming $R_{\rm GMC}\sim 100$~pc and $\Sigma_{\rm GMC}\sim 1000\, M_\odot/{\rm pc}^2$;][]{Grudic2023}. However we also test an exponential decay SFH prescription in SED fitting on $1b$ to evaluate its impact. To conduct this, we use \texttt{Bagpipes} \citep{Carnall2018}, as the SED fitting approach used in this work only includes the instant burst SFH. Indeed the age and metallicity binning approach of \texttt{Bagpipes} is likely more suitable for extended SFHs (e.g., not a delta function). We choose our parameter priors to match those described in section Sec.~\ref{sec:meth}, noting that the covering factor is not available in \texttt{Bagpipes} which effectively assumes a covering factor unity. We allow $\tau$ to freely vary from 1~Myr to 1~Gyr which lends much greater flexibility to the SED compared to a pure SSP, however we note that this can permit star formation timescales which may not necessarily reflect observations and simulations.

We find across all three SSPs that the `free $\tau$' inference agrees well with the instantaneous SFH prescriptions, detailed in Table~\ref{tab:sfh}. In each, $\tau$ converges to $<$~few Myr, effectively becoming a `short burst' which closely matches a true SSP. We also explore cases where nebular emission is removed, which returns nearly identical results for \texttt{BPASS} and \texttt{FSPS}, however the BC03 fit returns a young ($\sim6\,$Myr) solution with much higher metallicities $\sim 20\%$ solar. We ultimately disfavor this solution given the lack of detectable nebular emission, whereas nebular emission is readily detected in other young massive clusters within the Sunrise \citep{Vanzella2023}. If such a solution is true and this young cluster is somehow void of gas, {\it JWST} medium resolution spectra could reveal stellar wind features, namely the \ion{C}{4}$\lambda1550$ P-Cygni feature, which would be strong at that age and metallicity.

\begin{deluxetable*}{ccccccccc}
\tabletypesize{\footnotesize}
\tablecaption{Free $\tau$ SFH SED Parameters}
\label{tab:sfh}
\tablecolumns{9}
\tablehead{
\colhead{\bf Image} &  \colhead{\bf SSP} &  \colhead{\bf $\log(t_{\rm age}/{\rm yr})$} &  \colhead{\bf $\log(\mu M/M_\odot)$}  & \colhead{\bf $\log(Z/Z_\odot)$} & \colhead{\bf $\log(\tau/{\rm Gyr})$} & \colhead{$A_{\rm V}$~(mag)} & \colhead{$\delta$} & \colhead{$B$}}
\startdata
Im 1b & BC03 & $7.50^{+0.04}_{-0.01}$ & $9.19^{+0.02}_{-0.01}$ & $-2.10^{+0.06}_{-0.04}$ & $-2.56^{+0.18}_{-0.24}$ & $0.06^{+0.03}_{-0.02}$ & $-1.28^{+0.25}_{-0.24}$ & $2.05^{+1.56}_{-0.24}$ \\
 & \texttt{BPASS} & $7.69^{+0.05}_{-0.08}$ & $9.30^{+0.03}_{-0.04}$ & $-2.10^{+0.17}_{-0.12}$ & $-2.52^{+0.32}_{-0.29}$ & $0.05^{+0.04}_{-0.02}$ & $-1.13^{+0.29}_{-0.27}$ & $1.77^{+1.60}_{-1.23}$ \\
 & \texttt{FSPS} & $8.20^{+0.03}_{-0.05}$ & $9.67^{+0.02}_{-0.04}$ & $-2.27^{+0.10}_{-0.02}$ & $-2.23^{+0.42}_{-0.54}$ & $0.03^{+0.05}_{-0.02}$ & $0.29^{+0.35}_{-0.58}$ & $2.21^{+1.74}_{-1.37}$ \\
\enddata
\tablecomments{Physical parameters for Image $1b$ using \texttt{BAGPIPES} with the BC03, \texttt{BPASS}, and \texttt{FSPS} SSP libraries and an exponential decay SFH. We list the object, the SSP library, the population age, the magnified stellar mass ($\mu$ is the flux magnification factor due to lensing), the stellar metallicity, the characteristic timescale ($\tau$), visual band extinction, reddening law slope following \cite{Salim2018}, and reddening law 2175~\AA bump strength following \cite{Salim2018}.}
\end{deluxetable*}

\section{Testing Metallicity Systematics} \label{Testmet}
The precision achieved on the metallicity in each SSP is impressive, at the $<0.2$~dex level for the higher SNR $1a,b$. SSP fitting of the rest-UV through Optical integrated cluster continuum at this spectral resolution for these intermediate ages is somewhat unexplored, with the majority of works in both the local universe and at high redshift relying solely on photometry \citep[e.g.,][]{Wofford2016} or lacking the rest-FUV . Without the availability of metal-line absorption features, our fitting implies that the continuum alone provides non-negligible leverage to break well-known age-dust-metallicity degeneracies. We develop three tests to further probe the effectiveness of the continuum for constraining the physical parameters of the stellar population in this intermediate age, metal-poor regime.

\begin{figure}[!htb]
\centering
\includegraphics[width=\columnwidth]{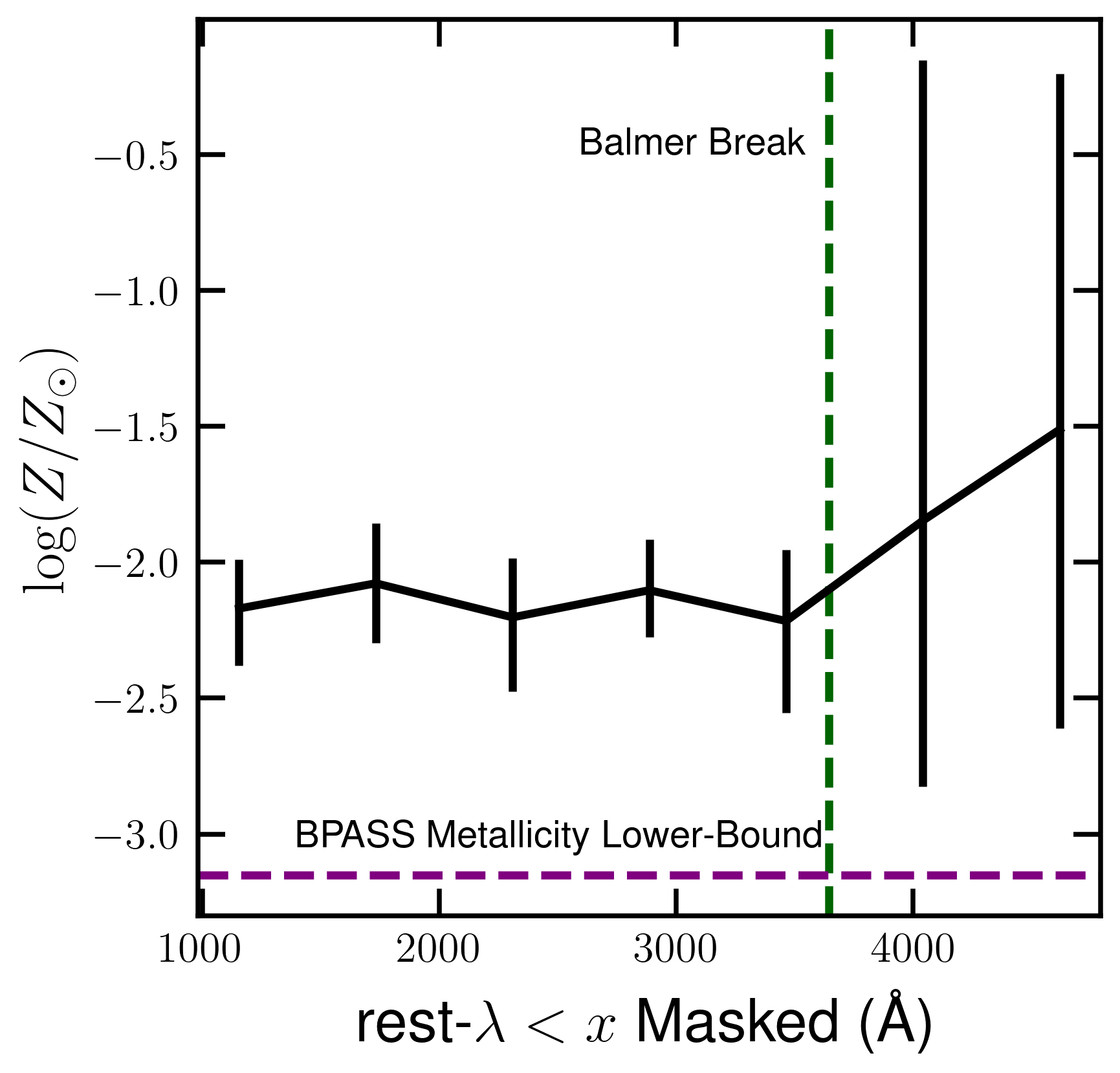}
\includegraphics[width=\columnwidth]{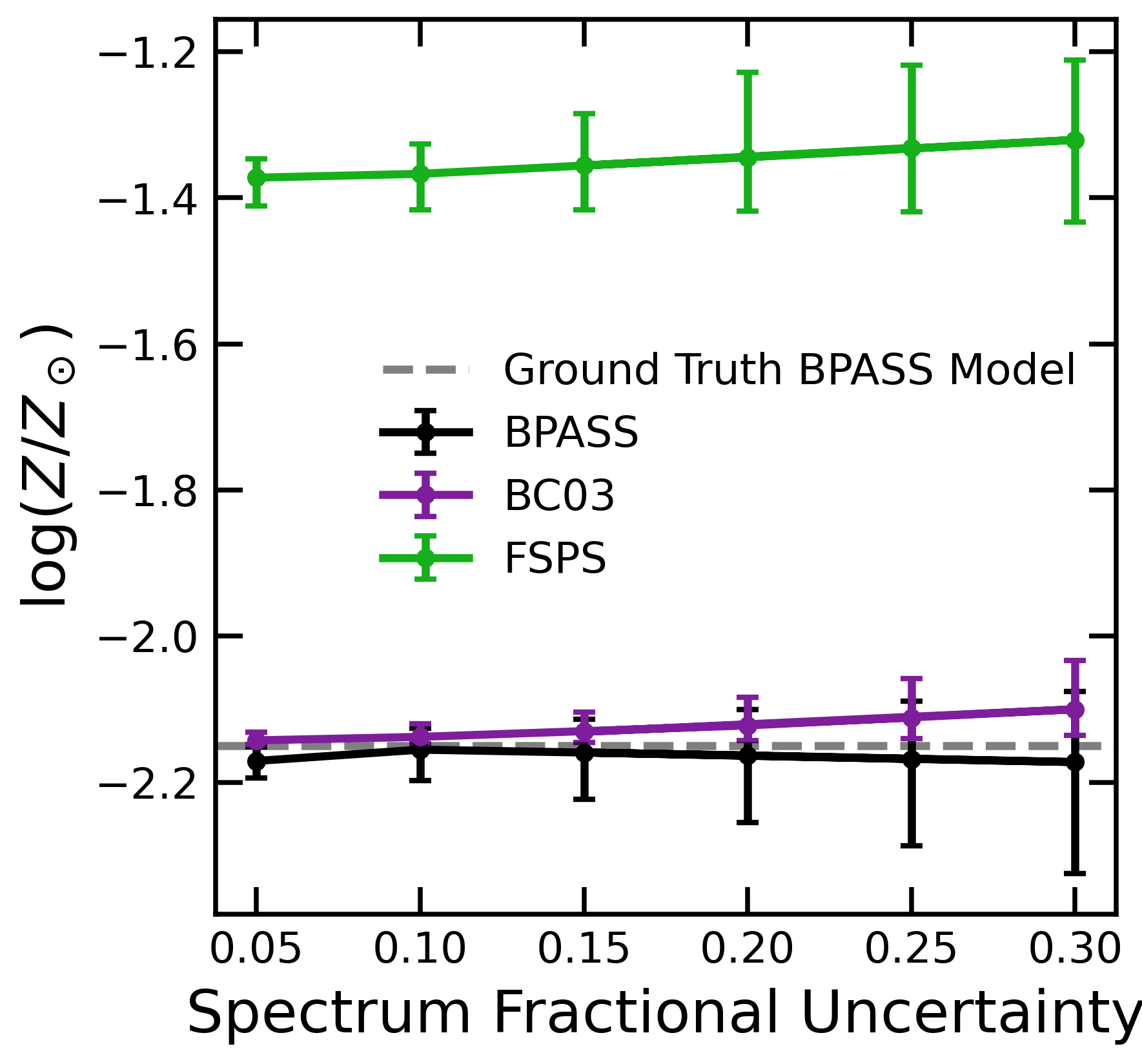}
\label{fig:met_tests}
\caption{ {\bf Top:} Metallicites inferred from the $1b$ NIRSpec PRISM spectrum using the \texttt{BPASS} SSPs (black error bars) with an iterative masking of wavelengths from the rest-FUV to the rest-Optical. Masking the FUV through NUV does not appear to significantly affect the inferred metallicity, however masking beyond the Balmer break (green dashed line) which is clearly detected in the NIRSpec spectrum (Fig.~\ref{fig:spec}) results in the metallicity becoming effectively unconstrained.
{\bf Bottom:} Metallicities inferred via the \texttt{BPASS} (black), BC03 (purple) and \texttt{FSPS} (green) SSPs from a mock \texttt{BPASS} model of $\log(Z/Z_\odot)=-2.15$, $\log(t_{\rm age}/{\rm yr})=7.7$ and $A_V=0$ with increasing uniform fractional Gaussian noise applied. While \texttt{BPASS} and BC03 recover the ground truth (grey dashed line), \texttt{FSPS} exhibits an offset of $\sim0.8$~dex towards higher metallicities.
}
\end{figure}

We first use the spectrum of $1b$ and iteratively mask out portions of the spectrum from the FUV to the optical, at each step performing fitting with the \texttt{BPASS} SSP to infer the metallicity (Fig.~\ref{fig:met_tests}; top). Surprisingly, removal of both the FUV and NUV ($<3800$\AA~ rest frame) is only modestly impactful, softening the metallicity constraints to $\sim0.5$~dex. It is only after further masking beyond the Balmer break to $<4800$\AA~(roughly SDSS $g$-band) which finally eliminates the metallicity constraints, producing an approximately uniform posterior across a range of metallicities $0.005\leq Z/Z_\odot\lesssim1$.

We identify that this is due in part to the formation of younger age solutions $\lesssim10$~Myr with low covering factors $x<10\%$ which favor all metallicities roughly uniformly. In practice this could arguably be disfavored apriori by the small covering which would be surprising for such a young cluster, however we retain these solutions for the sake of thoroughness. This exercise confirms that the clear Balmer break detected in Earendel and $1b$ is a defining feature in our inference, helping to break age-dust-metallicity degeneracies at these intermediate ages.

\begin{figure}[!htb]
\centering
\includegraphics[width=\columnwidth]{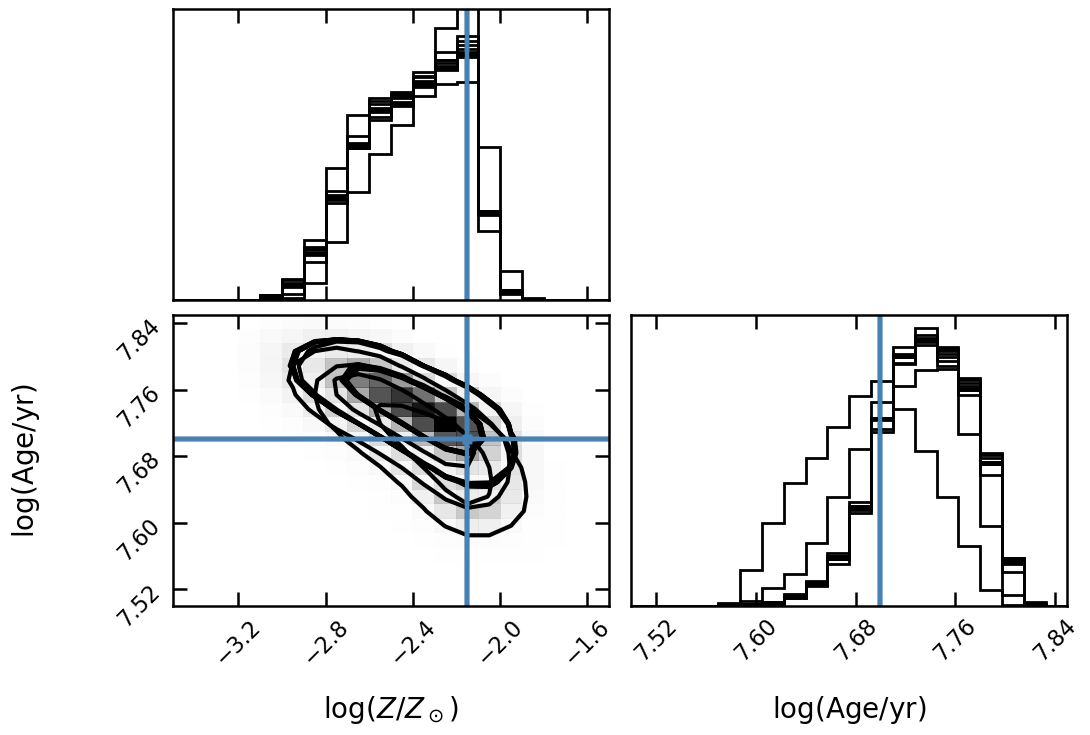}
\label{fig:noisymodels}
\caption{Metallicites and ages posteriors from fitting the \texttt{BPASS} SSPs to 10 mock spectra constructed using a \texttt{BPASS} model of $\log(Z/Z_\odot)=-2.15$, $\log(t_{\rm age}/{\rm yr})=7.7$ and $A_V=0$ with noise realizations applied which approximate the wavelength-dependent noise of the $1b$ NIRSpec PRISM spectrum. 2D contours enclose 50\% and 90\% of the posterior sample. The ground truth (blue lines) is generally recovered to within $1\sigma$, however the metallicity exhibits a noticeable bias towards lower metallicities. 
}
\end{figure}
We construct a second test to ascertain how the constraints are affected by the noise level. We create a mock \texttt{BPASS} spectrum of age 50~Myr, metallicity $\log(Z/Z\odot)=-2.15$, and $A_V=0$ (approximately the $1b$ best fit) without any nebular emission (e.g., $x=0$). We convolve to the PRISM resolution and iteratively implement increasing fractional uniform Gaussian noise scalings from $\sigma_{\rm noise} = 5-30\%$ of the model flux. For each noise scaling, we perform fitting using each of the three SSPs and evaluate the inferred metallicity, shown in Fig.~\ref{fig:met_tests} (bottom). Each model shows general agreement with itself across the different noise levels, and the BC03 and \texttt{BPASS} fits generally agree with the ground truth \texttt{BPASS} model metallicity. \texttt{FSPS}, however, shows a $\sim 0.8$~dex offset toward higher metallicities. This disagreement is not seen in the fits to Earendel or $1b$, and is perhaps somewhat surprising given that both \texttt{BPASS} and \texttt{FSPS} rely on the C3K spectral library in this metallicity regime. Inspection of the posteriors reveals that the \texttt{FSPS} fits converge to a slightly lower $\sim30$~Myr age solution; extrapolating age-metallicity degeneracy out to older ages could indeed yield lower metallicities, however these solutions are highly disfavored even in the noisiest models.

Finally, we conduct a third test to evaluate whether the results are a product of a random noise realization. To do so we once again construct a mock \texttt{BPASS} model as described above. To apply noise, we take the fractional uncertainties of the $1b$ spectrum, smooth it with a median filter to yield the approximate $1\sigma$ noise level, and apply random realizations of Gaussian noise to the mock spectrum following these uncertainties to create 10 mock spectra. We then fit each of these spectra following our SED fitting procedure with the \texttt{BPASS} models to check if the metallicity is generally well-recovered. 

As seen in Fig.~\ref{fig:noisymodels}, we find the inferences on these models both generally agree with one another, as well as the ground truth to within $1\sigma$. Somewhat concerningly, we do observe a systematic bias towards lower metallicities across all of these fits of $\sim 0.1-0.2$~dex. As these do not appear in the prior test which uses uniform fraction noise, this could indicate that this is actually a result of the wavelength dependent noise in our $1b$ spectrum. It is possible that correlated or wavelength-dependent noise could be biasing the fit, which may also be present in Earendel and $1a$ (in fact they must, as $1a$ and $1b$ yield similar results in our analysis). However we find that changing the size of the median filter can change both the strength and direction of the bias, cluing that this may instead be a systematic introduced by our noise implementation. Taking this in combination with the magnitude of the biases being comparable to the uncertainties of our inferred metallicities, we conclude that this bias, if real, would have only a modest impact on this work given the current precision of our constraints.

\end{appendix}
\end{document}